\documentclass[twocolumn,tighten,times,twocolappendix]{aastex631}

\newcommand{\zmax}{\strut z^{max}_{10^4\hskip 0.3pt \mathrm{K}}}

\newcommand{\soutdis}[1]{}

\received{July 13, 2022}
\revised{July 28, 2022}
\accepted{July 30, 2022}
\submitjournal{ApJL}

\shorttitle{A 2D model for CBPs}
\shortauthors{N\'obrega-Siverio and Moreno-Insertis}

\begin{document}

\title{\Large{A 2D Model for Coronal Bright Points:\\ 
Association with Spicules, UV bursts, Surges and EUV Coronal Jets}}

\correspondingauthor{D. N\'obrega-Siverio}
\email{dnobrega@iac.es}

\author[0000-0002-7788-6482]{D. N\'obrega-Siverio}
\affiliation{Instituto de Astrof\'isica de Canarias, 
            E-38205 La Laguna, Tenerife, Spain}
\affiliation{Universidad de La Laguna, Dept. Astrof\'isica, 
            E-38206 La Laguna, Tenerife, Spain}
\affiliation{Rosseland Centre for Solar Physics, University of Oslo, 
            PO Box 1029 Blindern, 0315 Oslo, Norway}
\affiliation{Institute of Theoretical Astrophysics, University of Oslo, 
            PO Box 1029 Blindern, 0315 Oslo, Norway}

\author{F. Moreno-Insertis}
\affiliation{Instituto de Astrof\'isica de Canarias, 
            E-38205 La Laguna, Tenerife, Spain}
\affiliation{Universidad de La Laguna, Dept. Astrof\'isica, 
            E-38206 La Laguna, Tenerife, Spain}

\begin{abstract}
Coronal Bright Points (CBPs) are ubiquitous structures in the solar
atmosphere composed of hot small-scale loops
observed in EUV or X-Rays in the quiet Sun and coronal holes. They are
key elements to understand the heating of the corona; nonetheless, basic questions 
regarding their heating mechanisms, the chromosphere underneath, or the effects of 
flux emergence in these structures remain open.
We have used the Bifrost code to carry out a
2D experiment in which a coronal-hole magnetic  nullpoint
configuration evolves  perturbed by realistic granulation. To compare with 
observations,  synthetic SDO/AIA, Solar Orbiter EUI-HRI, and IRIS images 
have been computed.
The experiment shows the self-consistent creation of a CBP 
through the action of the stochastic granular motions alone,
  mediated by  magnetic reconnection in the corona.  The reconnection is intermittent and
  oscillatory, and it leads to coronal and transition-region temperature loops
that are identifiable in our EUV/UV observables.
During the CBP lifetime, convergence and cancellation at the surface of its underlying  
opposite polarities takes place.  The chromosphere below 
  the CBP shows a number of peculiar features concerning its
  density and the spicules in it. 
The final stage of the CBP is eruptive: magnetic flux emergence at
the granular
scale disrupts the CBP topology, leading to different ejections, such as UV
bursts, surges, and EUV coronal jets.
Apart from explaining observed CBP features, our results pave the way
for further studies combining simulations and coordinated
observations in different atmospheric  layers.
\end{abstract}


%
%
\section{INTRODUCTION}\label{sec:intro}
Coronal Bright Points (CBPs) are a fundamental building block in the solar
atmosphere. Scattered over the whole disc, CBPs consist of sets of coronal
loops linking opposite polarity magnetic patches in regions of,
  typically, from $4$ to $43$ Mm transverse size, with heights ranging from
  $5$ to $10$ Mm \citep[see the review by][]{Madjarska:2019}.  One of
their most striking features is the sustained  emission, for periods
  of several hours up to a few days, of large amounts of energy, which lend
them their enhanced extreme-ultraviolet (EUV) and X-ray signatures \citep[e.g.,][]{Golub_etal:1974}.

A significant fraction of the CBPs are observationally found to be formed
as a consequence of chance encounters of opposite magnetic polarities 
at the surface \citep[e.g.,][]{Harvey:1985,Webb_etal:1993,Mou_etal:2018}. 
First theoretical explanations about this mechanism came in the 1990's through 
analytical models under the name of Converging Flux Models
\citep{Priest_etal:1994,Parnell_Priest:1995}.  There, the approaching motion in the photosphere
of two opposite polarities that are surmounted by a nullpoint leads to
reconnection and heating of coronal loops.  Since then, this idea has been
extended and studied using magnetohydrodynamics (MHD) experiments
\citep[e.g.,][]{Dreher_etal:1997,Longcope:1998,Galsgaard_etal:2000,von-Rekowski_etal:2006a,Santos_Buchner:2007,
  Javadi_etal:2011,Wyper_etal:2018b,Priest_etal:2018,
  Syntelis_etal:2019}. However, the available CBP models are
idealized, i.e., they rely on ad-hoc driving mechanisms, which do not reflect
the stochastic granular motions, they lack radiation transfer to
model  the lower layers of the atmosphere, and/or
miss optically thin losses and/or thermal conduction to properly
capture the CBP thermodynamics.

To understand the physics of CBPs, 
realistic numerical experiments are needed to address important open questions such
as (a) the CBP energization, focusing on whether the granulation is enough to
drive and sustain the reconnection at coronal heights to
explain the CBP lifetimes; (b) the role of magnetic flux emergence, specially
at the granular scale, 
to know whether it can disrupt the CBP topology and originate an eruption; 
and (c) the chromosphere underneath a
CBP, with the aim of unraveling the unexplored impact of CBPs on the spicular
activity and vice-versa.

In this letter, we model a CBP through the evolution of an initial 
fan-spine nullpoint configuration. The choice of this configuration
is because CBPs often appear above photospheric regions with a parasitic
magnetic polarity embedded in a network predominantly of the opposite 
polarity, which typically leads to a nullpoint structure in the corona \citep{Zhang_etal:2012,Galsgaard_etal:2017,Madjarska_etal:2021}.  
The 2D experiment is carried out with the Bifrost code \citep{Gudiksen_etal:2011}, 
which self-consistently couples the different layers of the solar atmosphere and 
incorporates several physical mechanisms not included in CBP modeling in the past. 
To provide a direct link to observations, we calculate
synthetic EUV images for SDO/AIA \citep[][]{Pesnell_etal:2012,Lemen_etal:2012},
and Solar Orbiter (SO)/EUI-HRI \citep[][]{Muller_etal:2020,Rochus_etal:2020}, and
UV images for IRIS \citep[]{De-Pontieu_etal:2014}.

\begin{figure}[!ht]
        \centering
        \includegraphics[width=0.5\textwidth]{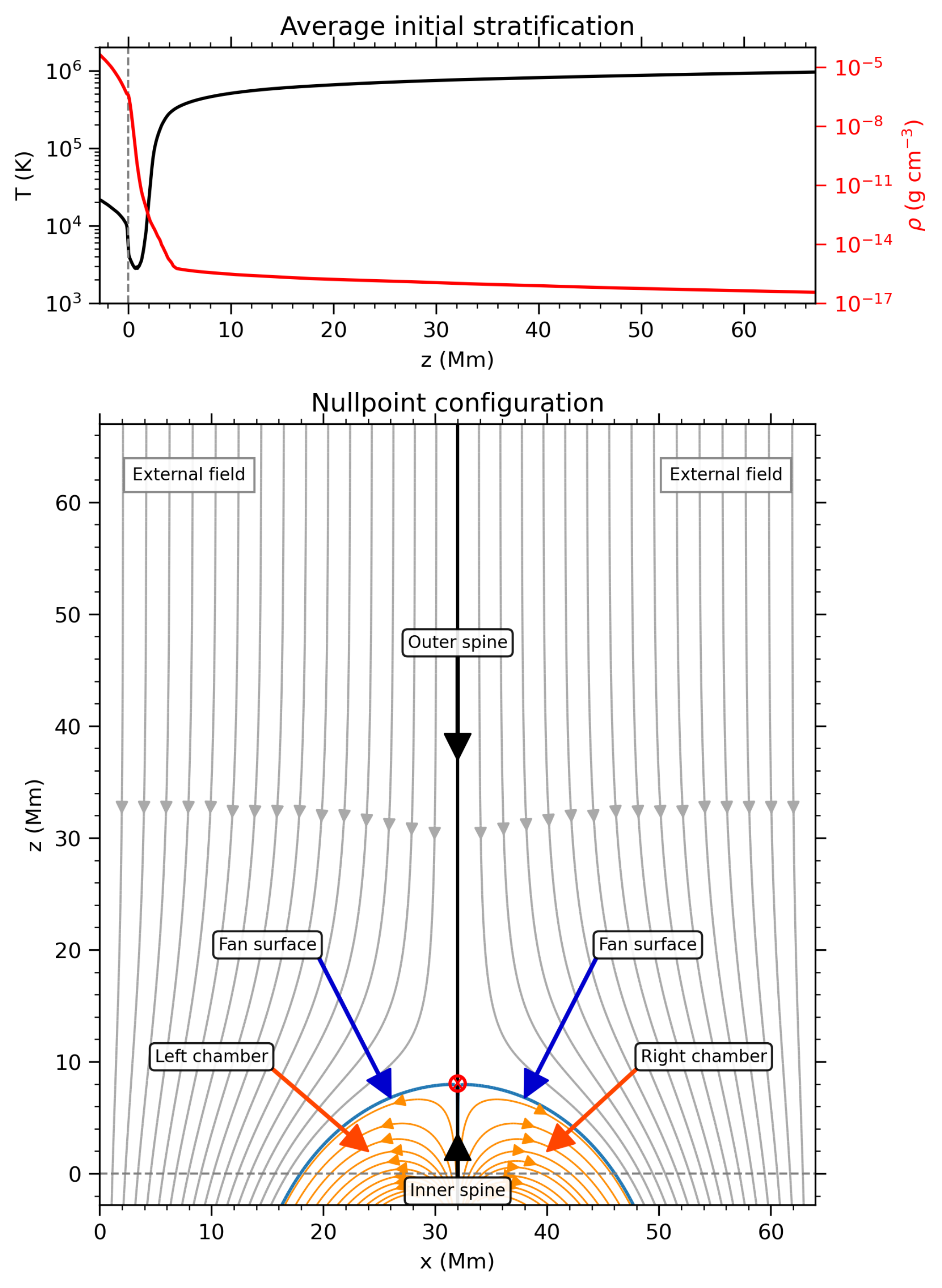}
        \caption{
        Initial condition.
        Top: Average stratification for the temperature and density.
        Bottom: Imposed magnetic field configuration  
        indicating some of its major features following
        the terminology used in 3D nullpoints \citep{Priest_Titov:1996}. 
        Red circle: region with 
        $B < 1$~G centered on the nullpoint,
        $(x_0, z_0) = (32, 8)$~Mm. Grey-dashed
        line: solar surface at $z=0$.}  
        \label{fig:01}
\end{figure} 

%
%
\section{METHODS}\label{sec:methods}

\subsection{Code}\label{sec:code}

The experiment has been performed using Bifrost, a radiation-MHD code 
for stellar atmosphere simulations \citep{Gudiksen_etal:2011}. 
The code includes radiative transfer from the photosphere to
the corona; the main losses in the chromosphere by neutral
hydrogen, singly-ionized calcium and magnesium; thermal conduction
along the magnetic field lines; optically thin cooling; and an 
equation of state with the 16 most important atomic elements in the Sun.

\subsection{Initial Condition}\label{sec:initial_condition}

\subsubsection{Background Stratification}\label{sec:background}
The initial condition has been constructed using as background
a preexisting 2D numerical simulation 
with statistically stationary convection in the photosphere and below. It 
encompasses from the uppermost layers of the solar interior to the corona
($-2.8$~Mm $\leq z \leq 67.0$~Mm, $z=0$ being the solar surface).
 The horizontal extent is $0.0$~Mm $\leq x
 \leq 64.0$~Mm.  The grid is uniform with $4096\times4096$ cells,
   yielding a very high spatial resolution of $\Delta x=15.6 $~km and
 $\Delta z=17.0 $~km. The boundary conditions are the same as for
 the experiment by \cite{Nobrega-Siverio_etal:2016}.  The top panel of
 Figure~\ref{fig:01} shows the horizontal averages of temperature and 
 density of the background stratification.

\subsubsection{Nullpoint Configuration}\label{sec:nullpoint}
Over the previous snapshot, we have imposed a
potential magnetic nullpoint configuration as shown in the bottom
panel of Figure \ref{fig:01}. The potential field was calculated 
from a prescribed distribution at the bottom boundary, $z=-2.8$~Mm.
The
  nullpoint is located at $(x_0, z_0) = (32, 8)$~Mm and the field
  asymptotically becomes vertical in height with $B_z=-10$~G,
  mimicking a coronal hole structure.
 The photospheric field
  contains a positive parasitic
  polarity at the center on a negative background. At $z=0$, the total positive flux and
  maximum vertical field strength are $\Phi^{+}=2.2\times10^{10}$~G~cm and
$B_z=41.3$~G, respectively; the parasitic polarity covers 9.7~Mm;
 and the fan surface extends for 28.1~Mm. We will refer to the
closed-loop domains on either side of the inner spine as \textit{chambers}.

\begin{figure*}[!ht]
        \centering
        \includegraphics[width=1.00\textwidth]{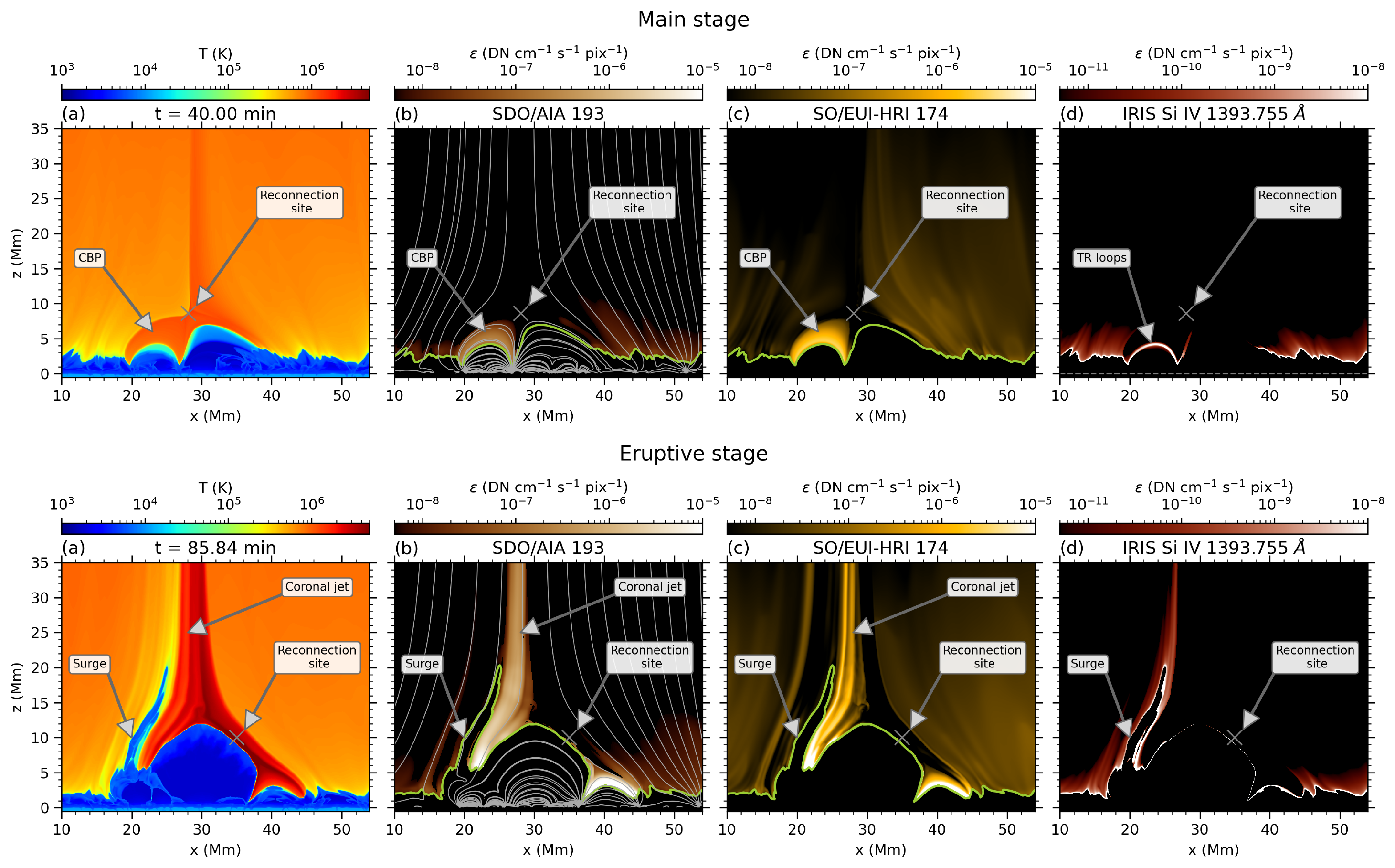}
        \caption{
        Experiment overview. 
        Top: Main stage illustrating a CBP at $t=40.00$ min. 
        Bottom: Eruptive stage showing a surge and coronal jet at $t=85.84$~min. 
        (a) Temperature.
        (b) Synthetic SDO/AIA~193 with superimposed magnetic field lines.
        (c) SO/EUI-HRI~174. 
        (d) IRIS \ion{Si}{4} 1393.755~\AA.
        For the synthesis details, check Appendix~\ref{app:synthetic}.
        Olive line on Panels (b) and (c): $T=10^5$~K isocontour.
        The associated movie comprises the whole experiment evolution
        from $t=0$ to $t=85.84$~min, illustrating the development of both stages.
        \\
        (An animation of this figure is available.)
        } 
        \label{fig:02}
\end{figure*} 

%
%
\section{RESULTS} \label{sec:results}

\subsection{Overview of the Experiment}\label{sec:overview}
Figure~\ref{fig:02} and associated movie provide an overview of the system
evolution using temperature maps and synthetic observables 
for coronal and transition region (TR) temperatures (see Appendix \ref{app:synthetic}).
In the experiment, the granulation quickly distorts the imposed
magnetic field and reconnection is
triggered as a consequence of perturbations at the nullpoint. We distinguish
two well-defined phases: main stage and eruptive stage, summarized in the
following and described in Sections~\ref{sec:main_stage} and
\ref{sec:emergence}, respectively. 

The main stage (from $t=0$ to $t\approx 65$~min) covers the appearance of
post-reconnection hot loops that leads to a CBP. For instance, 
at $t=40$~min (top row of
Figure~\ref{fig:02}), our CBP is discernible as a set of hot coronal loops with
enhanced SDO/AIA~193 and SO/EUI-HRI~174 emission above TR 
temperature loops visible in IRIS \ion{Si}{4}. This
matches the SDO/IRIS observations by \cite{Kayshap_Dwivedi:2017}, where CBPs
are inferred to be composed of hot loops overlying
cooler smaller ones. Our CBP is
 found in the left chamber, indicating that there
  is a preferred reconnection direction,  and it gets more compact with
time before vanishing, as reported in observations
\cite[]{Mou_etal:2018}.

In the eruptive stage, the CBP comes to an end. Around $t=67$~min in the animation, 
a first hot ejection results from reconnection between
emerging plasma at granular scale and the magnetic field of the right
chamber, resembling the UV burst described by \cite{Hansteen_etal:2019}.
More episodes of flux emergence and reconnection take place
subsequently, dramatically destabilizing
the system and producing more ejections. As an example, at $t=85.84$~min
(bottom row of Figure \ref{fig:02}) a large and broad hot coronal jet
is seen next to a cool surge, reminiscent of previous
results by
\cite{Yokoyama_Shibata:1996,Moreno-Insertis_Galsgaard:2013,Nobrega-Siverio_etal:2016}.
The synthesis clearly reflects the high spatial resolution of SO/EUI-HRI to study coronal jets 
and shows that the surroundings of the surge have significant \ion{Si}{4} emission, 
similarly to observations
\cite[e.g.,][]{Nobrega-Siverio_etal:2017,Guglielmino_etal:2019}.


\begin{figure*}[!ht]
        \centering
        \includegraphics[width=1.00\textwidth]{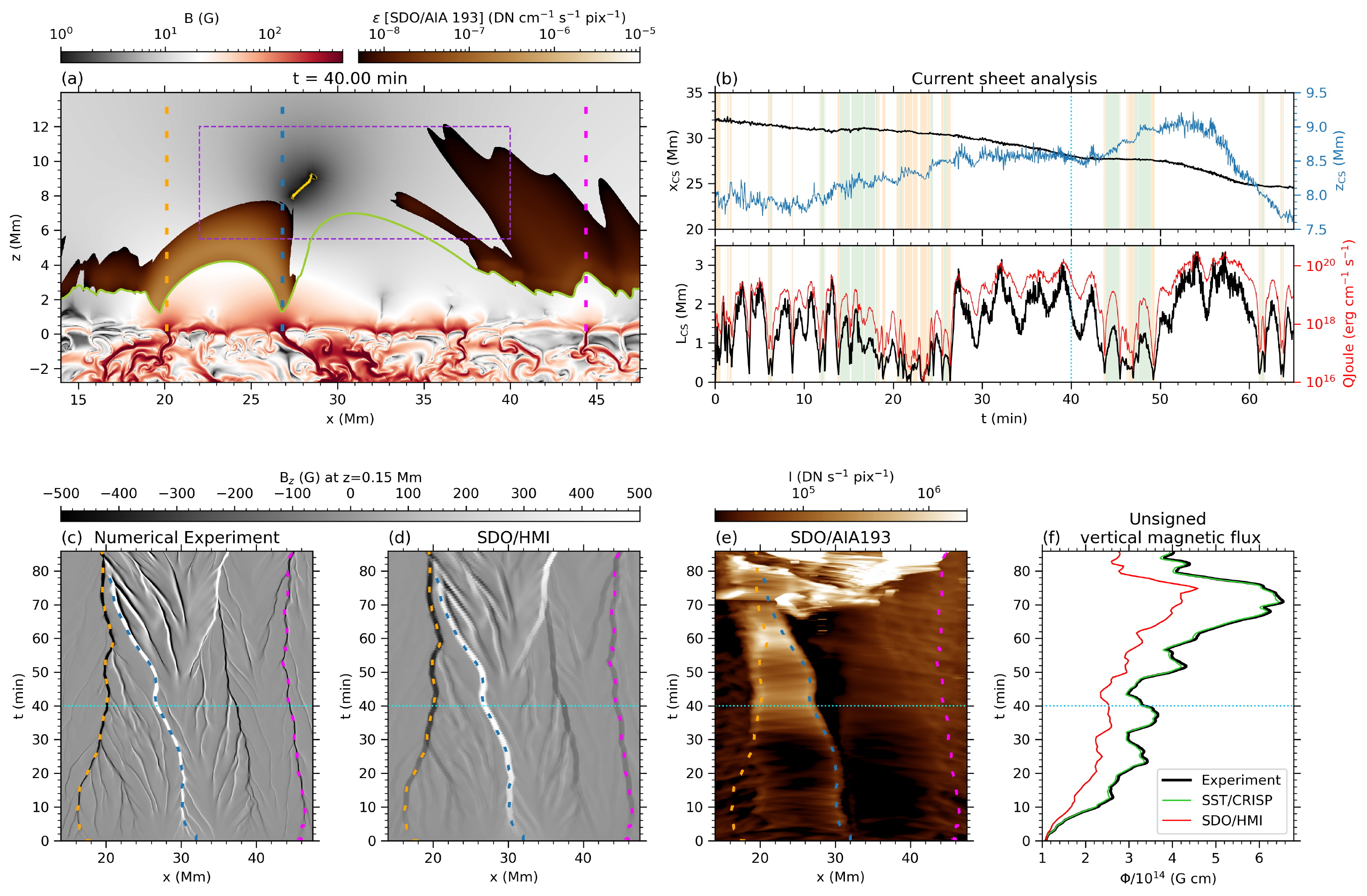}
        \caption{Magnetic reconnection and flux emergence.
        (a) $B$ at $t=40.00$~min with superimposed AIA 193 response 
        only for the values contained in the corresponding color scale.
        Olive line: $T=10^5$~K isocontour. Purple rectangle: region for 
        the current sheet (CS) analysis. Yellow dots: CS region given by
        $L_B \leq 100$~km, as defined in Appendix~\ref{app:current_sheet}. 
        (b) CS analysis. Top: CS center coordinates, 
        $x_{\mathrm{CS}}$ and $z_{\mathrm{CS}}$.
        Bottom: CS length, $L_{\mathrm{CS}}$, and
        Joule heating, $Q_{\mathrm{Joule}}$.
        When the CS is elongated, positive/negative angles of reconnection are indicated with white/green areas; otherwise, using orange.
        (c) Space-time map for $B_z$ at $z=0.15$ Mm. 
        (d) View of (c) with the HMI resolution. 
        (e) AIA 193 intensity integrated along the vertical direction.
        (f) Unsigned vertical magnetic flux for the 
        experiment and corresponding ones for SST/CRISP and HMI resolutions measured at $z=0.15$~Mm.
        In the figure, colored dashed lines follow the polarities related to the
        left side of the fan surface (orange), the inner spine (blue), and the right side of the fan surface (magenta). 
        The cyan-dotted lines indicates the time shown in Panel (a).
        The associated movie shows the evolution of the magnetic field and SDO/AIA 193 response
        from $t=0$ to $t=85.84$~min.\\
        (An animation of this figure is available.)}
        \label{fig:03}
\end{figure*} 

\subsection{Main Stage}\label{sec:main_stage}

\subsubsection{Magnetic Reconnection and Association with
CBP Features}\label{sec:reconnection}

To illustrate the crucial role of coronal reconnection for the CBP
during the main stage,  
Panel (a) of Figure \ref{fig:03} contains the magnetic field strength 
with the SDO/AIA 193 response superimposed. Soon after starting 
the experiment (see animation), a current sheet (CS) is formed (yellow dots in the panel). 
Focusing on the region delimited by the purple rectangle, we distinguish three clear patterns: 
(a) the reconnection site slowly drifts to the left; (b) 
the reconnection and associated heating behaves in a bursty way; and (c) 
the reconnection is oscillatory.

The reconnection-site's displacement is shown in the top
frame of Panel (b).
The horizontal position of the CS center, $x_{\mathrm{CS}}$ (black),
moves 7.4 Mm to the left in 65 minutes, while its vertical position, 
$z_{\mathrm{CS}}$ (blue), first moves from 8.0~Mm up to 9.2 Mm, to
descend later to 7.6 Mm. These results are akin to 
observations, where CBP nullpoints are inferred to rise, descend, or show both
types of behavior in addition to important
horizontal displacement \citep{Galsgaard_etal:2017}.

The bursty behavior is depicted in the bottom frame  of
Panel (b). The CS length, $L_{\mathrm{CS}}$, 
abruptly changes over times of minutes, reaching a maximum of $3.3$~Mm.
These fluctuations are well correlated
with the Joule heating released in the reconnection site, $Q_{\mathrm{Joule}}$: 
the Pearson correlation coefficient between both curves is 0.93. 
Note that when the diffusion region does not have an
elongated CS (orange background in the panel), the Joule heating is minimal.
This intermittent heating seems to be consistent with observed CBP emission
variations over timescales of minutes
\citep[see, e.g,][]{Habbal_Withbroe:1981,Ugarte-Urra_etal:2004,Kumar_etal:2011,Ning_Guo:2014,Chandrashekhar_Sarkar:2015,Gao_etal:2022}.

The elongated CS also changes its angle, $\theta_{\mathrm{CS}}$ (defined anticlockwise with respect to the x-axis),
several times, from, approximately,
$45$ to $-45$ degrees, indicating oscillatory reconnection. Panel (b) shows
 a white/green background for the intervals when $\theta_{\mathrm{CS}}$ is positive/negative.
 Most of the time, the angle is positive: the reconnection
inflows come from the right chamber and the upper-left part of the external
field, while the outflows are located in the left chamber and upper-right
part of the external field.  This explains why the left chamber is the one
showing the CBP, as hot post-reconnection loops are being deposited
predominantly in this region. The predominance seems to be associated with
the concentration near the inner spine of the structure, indicated with a blue-dashed line in Panel
(a), which mainly moves to the left (see animation). Oscillatory
reconnection is also found in CBP observations \citep{Zhang_etal:2014}.

\subsubsection{The Photospheric Magnetic Field} 
\label{sec:main_stage_photospheric}

Figure~\ref{fig:03} contains space-time magnetograms near the photosphere ($z=0.15$~Mm) with the actual
resolution of the simulation (Panel (c)) and reducing it to the level of the
SDO/HMI instrument (Panel (d), see Appendix \ref{app:synthetic}). To emphasize the CBP evolution, 
Panel (e) contains a space-time
diagram showing the synthesized intensity of SDO/AIA 193 integrated along
the vertical line-of-sight.  
Panels (a), (c) and (d) show
that the magnetic field around the photospheric basis is quickly
concentrated by the granular motions, leading to strong magnetic
patches at the solar surface.
In the figure, we have highlighted with colored dashed lines
the concentrations that have collected the
  field lines near the fan surface on each side (orange and magenta)
  and near the inner spine (blue). The magnetic field concentrations are buffeted
  and dragged by the granular motions while being substantially deformed in
the convection zone.  
In fact, the one related to the inner spine gets bent
several times underneath the surface and develops horizontal magnetized
structures (see animation of Panel (a) from $t\approx20$ to $30$~ min at
$z\approx-1$~Mm and $x$ between 30 and 32~Mm). Simultaneously with the convergence at the photosphere of the two
strong opposite concentrations (orange and blue lines), the
post-reconnection loops in the corona start to brighten up in the EUV bands
(see the superimposed SDO/AIA 193 response in Panel~(a) and the
  space-time diagram of Panel~(e))
 marking  the appearance
of our CBP at around $t=30$~min.   From observations, convergence is
frequently involved in CBP formation \citep[see, e.g.,][and references
  therein]{Mou_etal:2016,Mou_etal:2018}.  The convergence continues,
  and the CBP goes through a quiet phase until around $t=65$~min, when the
  eruptive phase starts, triggered by flux emergence, as explained in the
  next section.

\subsection{Eruptive Stage}\label{sec:emergence}

\subsubsection{Magnetic Flux Emergence at the Surface}
The strong and complex subphotospheric magnetic structures developing
  around the inner spine mentioned in the previous section become buoyant and rise. 
  They reach the surface to the right of the inner spine
from $t=42$~min onwards, leading to 
anomalous granulation and increasing the unsigned vertical magnetic flux. 
The enhanced magnetic pressure in the anomalous
granules further pushes the strong positive patch to the left, accelerating
the convergence of the two main opposite polarities of the CBP. 
Panel~(f) shows the total unsigned flux in the horizontal domain of Figure~\ref{fig:03},
i.e., 
\begin{equation}
    \Phi = \int_{x_0=14.0\, \mathrm{Mm}}^{x_f=47.5\, \mathrm{Mm}}{\left|B_z (z=0)\right|dx},
    \label{eq:flux}
\end{equation}
with the integral calculated with the $B_z$ 
distributions in the experiment (black curve) and in the reduced-resolution
counterparts for SST/CRISP (green) and SDO/HMI (red).  The unsigned flux
grows roughly by a factor two from $t=42$~min to $t=70.67$~min, when it
reaches its maximum.  Interestingly, in contrast to a high-resolution instrument like SST/CRISP, the
 reduced resolution of SDO/HMI would miss a significant fraction of the flux. 
From this time onwards, the unsigned magnetic flux at the surface decreases
while the two main polarities continue converging. This could be interpreted
as magnetic cancellation, a fact that is frequently observed at the end of 
CBPs \citep[e.g.,][]{Mou_etal:2016,Mou_etal:2018}.

\begin{figure*}[!ht]
        \centering
        \includegraphics[width=1.00\textwidth]{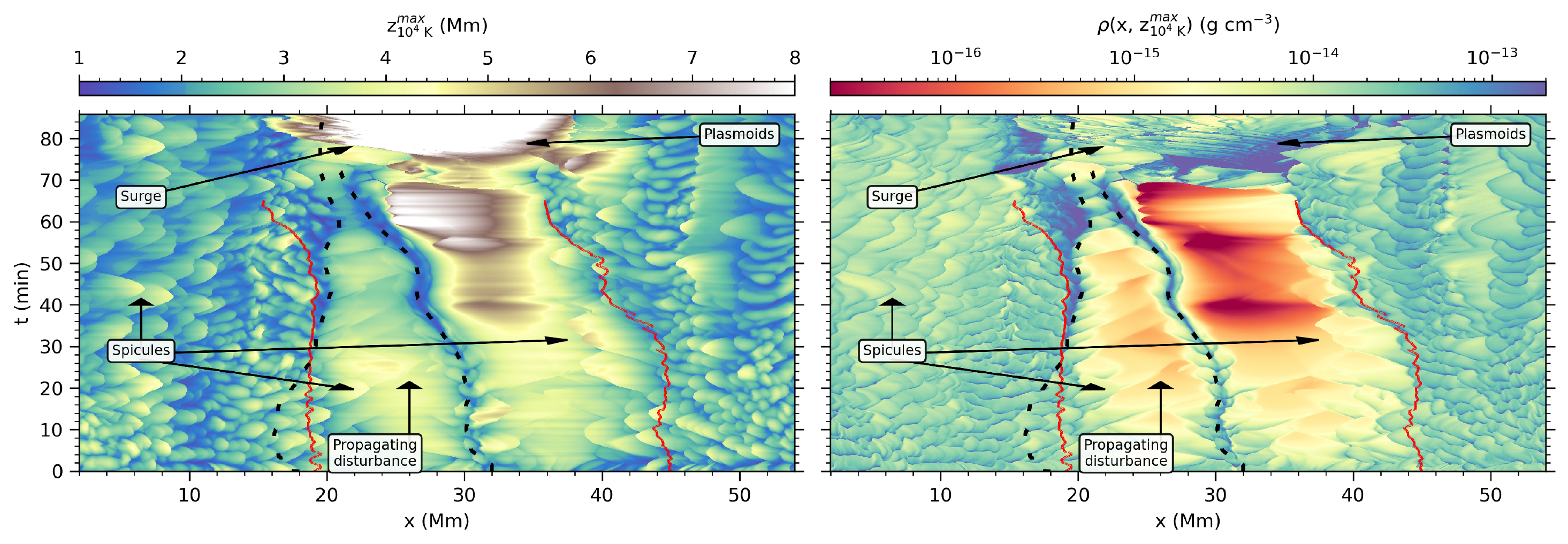}
        \caption{Chromospheric analysis.
        Left: Space-time maps for
        $\zmax(x)$. Right: Corresponding density $\strut \rho(x, \zmax)$. 
        Dashed lines: CBP's main polarities (see also Figure \ref{fig:03}). 
        Red-solid lines: fan surface at $z=1.5$~Mm in the main stage.}
        \label{fig:04}
\end{figure*} 

\subsubsection{Eruptive ejections}
In the atmosphere, the emerging field quickly expands, interacting with the
preexisting magnetic structure of the right chamber
and leading to eruptive behavior with different ejections. The first one
occurs at $t=67$~min, between $x=24$ and 30~Mm, and it resembles a UV burst
because of its enhanced TR
emission (see animation of Figure \ref{fig:02}).  The evolution of the system
becomes quite complex at this stage. The current sheet of the UV burst
interacts with the CBP current sheet, disrupting its magnetic configuration
and causing the end of our CBP (see the horizontal dark band at $t=71$~min in
Panel (d) of Figure~\ref{fig:03}).  Another ejection with EUV signatures
occurs right after, around $t=74$~min, developing an Eiffel tower shape,
followed by the ejection of the broad EUV jet (reaching up to 13~MK) 
and surge that are shown in the bottom row of Figure~\ref{fig:02}.

\subsection{The Chromosphere and Spicular Activity}\label{sec:spicules}
To analyze the chromosphere below the CBP, Figure \ref{fig:04}
illustrates space-time maps for $\zmax(x)$,
defined as the maximum height at which $T=10^4$~K for each $x$, and 
 the corresponding density $\strut \rho(x, \zmax)$ at that point. 
A major distinction is apparent between the regions inside and 
outside the fan surface. To clearly separate
  them, red-solid lines have been drawn at the horizontal position where the
  fan surface cuts the plane~$z=1.5$~Mm. Those outside have an interesting flake or {\it scale}-like
appearance. Each {\it scale} is flanked by a
quasi-parabolic trajectory signaling the rise and fall of individual
spicules: an example is indicated around $x=5$ Mm and $t\approx$~44 min.

During the main stage, mainly from $t=30$ to $t=65$~min, the strong field in
the CBP's opposite polarities (dashed tracks) lowers the height of the
  chromospheric level below $2$~Mm.  As these polarities converge, the
chromosphere underneath the CBP (i.e., in the left chamber)
 gradually moves
down, from $4$~Mm to 1~Mm, while the density roughly increases
by an order of magnitude there. 
This region also shows some
spicules, predominantly originated near the fan surface, with associated
propagating disturbances that cross the magnetic field loops of this
region (see example of both phenomena indicated within the left chamber).

The most striking characteristic of the right chamber is the enormous
rarefaction during the main stage resulting from the predominant reconnection
direction that extracts plasma from here (a fact also found in idealized 3D nullpoint simulations without flux emergence by Moreno-Insertis and Galsgaard, in preparation). In fact, the density can be 
so low as a few $10^{-17}$ g cm$^{-3}$: these are coronal densities with
cold temperatures. There are also spicular incursions in this
chamber mainly coming from regions around the fan surface (see example around
$x=39$~Mm and $t=$~32~min), but without associated propagating
disturbances.
In the eruptive stage, the leftmost part of the domain with $\zmax(x) > 8$~Mm corresponds 
to the ejection of the surge. Trajectories of multiple dense plasmoids expelled from the reconnection site are also visible with a fibril-like pattern.

%
%
\section{DISCUSSION} \label{sec:discussion}
In this letter we have shown that a wide class of CBPs may be obtained through magnetic reconnection in the corona driven by stochastic granular motions. Our numerical experiment has significant differences to previous CBP models: in contrast to, e.g., \citet{Priest_etal:1994,Priest_etal:2018} or \citet{Syntelis_etal:2019,Syntelis_Priest:2020}, the magnetic field topology consists in a nullpoint created by a parasitic polarity within a coronal hole environment; more importantly, the reconnection and creation of the CBP is self-consistently triggered by the convection that naturally occurs in the realistic framework provided by the Bifrost code. The latter feature also sets it apart from the idealized CBP coronal hole model of \citet{Wyper_etal:2018b}, in a which high-velocity horizontal driving at the photosphere is imposed to provide the energy released in the CBP.

Our experiment shows striking similarities to observed CBP features in spite of the simplified 2D configuration. 
For instance, the CBP is composed of loops at different temperatures,
with hotter loops overlying cooler smaller ones, and enhanced EUV/UV
emission akin to observations \citep{Kayshap_Dwivedi:2017}. The projected
length of these loops is around 8-10 Mm, which fits in the lower range of CBP
sizes \citep{Madjarska:2019}.  We can also reproduce other distinguishable
observational features such as the motion of the nullpoint
\citep{Galsgaard_etal:2017}, the convergence of the CBP footpoints
\citep[][and references therein]{Madjarska:2019}, the brightness
variations over periods of minutes
\citep{Habbal_Withbroe:1981,Ugarte-Urra_etal:2004,Kumar_etal:2011,Ning_Guo:2014,Chandrashekhar_Sarkar:2015,Gao_etal:2022}, as well as the oscillatory behavior \citep{Zhang_etal:2014}.

Magnetic flux emergence is crucial
for the formation of roughly half of the CBPs \citep{Mou_etal:2018}
and it can enhance the activity of already existing CBPs \citep{Madjarska_etal:2021}. In our model, flux emergence plays a
role for the final eruptive stage of the CBP: only a few granules were affected by the emergence, but the consequences for the CBP and the subsequent phenomena
are enormous. Observations with high-resolution magnetograms (e.g. 
by SST/CRISP) are needed to explore this relationship and to
discern whether hot/cool ejections in
the late stage of CBPs
\citep[e.g.,][]{Hong_etal:2014,Mou_etal:2018,Galsgaard_etal:2019,Madjarska_etal:2022}
may also follow, directly or indirectly, from small-scale flux emergence episodes.

The chromosphere underneath the CBP in the model shows a number of remarkable features. Spicules are
mainly originated from the fan surface (accompanied with propagating
disturbances), which could perturb the CBP
brightness (see, e.g., \citealp[][]{Madjarska_etal:2021} and Bose et al. in preparation).
The chromosphere related to the CBP footpoints is reached at low heights, 
while there is a chamber that
gets greatly rarefied because of the reconnection, reaching coronal densities
with chromospheric temperatures. These results can be potentially interesting
to unravel the chromospheric counterpart of CBPs and nullpoint configurations
in general and need to be explored using coordinated observations as well.

%
%
\begin{acknowledgements}
This research has been supported by the European Research Council through the
Synergy Grant number 810218 (``The Whole Sun'', ERC-2018-SyG) and by
the Spanish Ministry of Science, Innovation and Universities through project
PGC2018-095832-B-I00.  The authors 
acknowledge the computer resources at the MareNostrum supercomputing installation and the technical support provided by the Barcelona Supercomputing Center (BSC, RES-AECT-2021-1-0023), as well as the support by the International Space Science Institute (ISSI,
Berne) to the team
\textit{Unraveling surges: a joint perspective from numerical models,
  observations, and machine learning}.  The authors thank
Dr. Fr\'ed\'eric Auch\`ere for his help  
to compute the synthetic observables for SO/EUI-HRI and
Luc Rouppe van der Voort for illuminating conversations related
to SST/CRISP. The authors are also grateful to Drs. Klaus Galsgaard 
and Maria Madjarska as well as to the two referees for their
interesting comments and advice.
DNS acknowledges support by the
Research Council of Norway through its Centres of Excellence scheme, project
number 262622, and through grants of computing time from the Programme for
Supercomputing. 
\end{acknowledgements}

%
%
\appendix
\section{SYNTHETIC OBSERVABLES}\label{app:synthetic}
For the coronal and TR emissivity images, we assume statistical equilibrium and coronal abundances \citep{Feldman:1992}. Thus, the emissivity can be computed as
\begin{equation}
    \epsilon = n_H\ n_e\ G(T, n_e)\quad [\mathrm{erg}\, \mathrm{cm}^{-3}\, \mathrm{s}^{-1}\, \mathrm{sr}^{-1}],
    \label{eq:emiss}
\end{equation}
where $n_e$ is the electron number density, $n_H$ is the hydrogen number density, and $G(T, n_e)$ is the contribution function.


For the coronal 193 channel of the
Atmospheric Imaging Assembly \citep[AIA;][]{Lemen_etal:2012}
onboard the Solar Dynamics Observatory \citep[SDO;][]{Pesnell_etal:2012},
we generate a lookup table for $G(T, n_e)$ with {\tt aia\_get\_response.pro} from SSWIDL with the flags {\tt /temp} and {\tt /dn} varying the electron number density from $10^6$ to $10^{13}$ cm$^{-3}$. Once we compute the emissivity, we degrade it to the SDO/AIA spatial resolution, which is 1\farcs5 \citep{Lemen_etal:2012}. To obtain the intensity map shown in Panel (e) of Figure~\ref{fig:03}, we integrate the emissivity along the vertical line-of-sight assuming no absorption from cool and dense features like the surge. In addition, we degrade the space-time map to the SDO/AIA cadence, which is 12~s.

For the coronal 174 channel of
the Extreme Ultraviolet Imager  of the High Resolution Imager 
\citep[EUI-HRI;][]{Rochus_etal:2020} on Solar Orbiter \citep[SO;][]{Muller_etal:2020},
we use the contribution function privately provided by Dr. Fr\'ed\'eric Auch\`ere, member of the Solar Orbiter team, since, at the moment of writing this letter, the team is implementing the function in SSWIDL. The results are afterwards degraded to the EUI-HRI resolution, which corresponds to $100$~km pixel-size for perihelion observations \citep{Rochus_etal:2020}.

For the Interface Region Imaging Spectrograph \citep[IRIS;][]{De-Pontieu_etal:2014} 
TR \ion{Si}{4} 1393.755 \AA\ line, we create a lookup table employing {\tt ch\_synthetic.pro} from SSWIDL with the flag {\tt /goft} varying the electron number density from $10^6$ to $10^{13}$ cm$^{-3}$. The output of this routine is multiplied by the silicon abundance relative to hydrogen to obtain the contribution function. To transform CGS units to the IRIS count number, the emissivity is multiplied by $(A\, p\, w\, \lambda)/(k\, r^2\, h\, c)$, where $A = 2.2$ cm$^2$ pix$^{-1}$ is the effective area for wavelengths between $1389$ and $1407$~\AA, $p=0\farcs167$ is the spatial pixel size, $w=0\farcs33$ is the slit width, $\lambda=1393.755$~\AA\ is the wavelength of interest, $k=4$ is the number of photons per DN in the case of FUV spectra, $r=3600 \cdot 180/\pi$ is the conversion of arcsec to radians,  $h$ is the Planck constant, and  $c$ is the speed of light. Finally, we degrade the results to the IRIS spatial resolution of 0\farcs33 \citep[see][]{De-Pontieu_etal:2014}. Note that we have used statistical equilibrium as an assumption; however, nonequilibrium ionization effects are relevant for TR lines such as \ion{Si}{4} 1393.755 \AA. In fact, in dynamic phenomena like surges, statistical equilibrium underestimates the real population of \ion{Si}{4} ions, so the actual emissivity could be larger than the one shown in Figure~\ref{fig:02} \citep[see][and references therein for details]{Nobrega-Siverio_etal:2018}.


Regarding the SDO/HMI magnetogram of Panel (d) of Figure~\ref{fig:03}, we simply reduce the spatial/time resolution of the Panel (c) to the instrumental HMI values: 1\arcsec $\approx$ 726~km at 6173\AA\ and $45$~s of time cadence \citep{Scherrer_etal:2012}. The same approach is used for SST/CRISP \citep{Scharmer_etal:2008}, which has 0\farcs13 $\approx$~94 km at 6301\AA\ and $6$~s cadence, to obtain the unsigned vertical magnetic flux shown in Panel (f) of the figure. The chosen height to illustrate the magnetograms, $z=0.15$~Mm, is an approximation of the formation height of the \ion{Fe}{1} lines in which HMI and CRISP observe.

\section{CURRENT SHEET ANALYSIS}\label{app:current_sheet}
To analyze the current sheet (CS) behavior, we focus on the inverse characteristic length of the magnetic field 
\begin{equation}
	L_B^{-1} = \frac{ \left| \nabla \times
          \hbox{\textbf{\textit{B}}} \right| }{ \left|
          \hbox{\textbf{\textit{B}}} \right|}.
	\label{eq:lb}
\end{equation}
This quantity allows us to know where the abrupt changes of $B$ occur. 
The analysis is limited to the region within $22 \leq x \leq 40$~Mm and $5.5 \leq z \leq 12$~Mm
(purple rectangle in Panel (a) of Figure \ref{fig:03}), and for $0 \leq t \leq 65$~min to avoid having secondary current sheets related to the flux emergence episode described in Section \ref{sec:emergence}. In this region, we take all the grid points with $L_B^{-1}\geq 0.01$~km$^{-1}$, so $L_B \leq 100$~km (yellow dots in Panel (a) of Figure \ref{fig:03}), performing a linear fit to their spatial distribution.  The goodness of the fit (the $r^2$ parameter) tells us whether there is a collapsed/elongated CS or not. We have selected $r^2\geq 0.8$ as a criterion for a collapsed/elongated CS. 
Thus, we obtain the horizontal and vertical center position of the CS ($x_{\mathrm{CS}}$ and $z_{\mathrm{CS}}$, respectively); its length ($L_{\mathrm{CS}}$); 
and its angle ($\theta_{\mathrm{CS}}$), defined in the anticlockwise direction with respect to the $x$ axis, 
which is helpful to detect oscillatory reconnection.  
For example, at $t=17.67$~min, the linear fit is $z = -0.988\, x + 38.8$~Mm, with $L_{\mathrm{CS}}=1.08$~Mm, $r^2 = 0.946$, and $\theta_{\mathrm{CS}}=-44.6$ degrees, meaning that there is a collapsed CS whose inflows from the reconnection come from the left chamber and the upper-right part of the external field. 
In contrast, at $t=40.00$~min, the time shown in Figure
\ref{fig:03}, the linear fit is $z = 0.927\, x - 17.4$~Mm, with
$L_{\mathrm{CS}}=1.94$~Mm, $r^2 = 0.964$, and $\theta_{\mathrm{CS}}=42.8$
degrees, indicating that there is also an elongated CS but with reconnection
inflows occurring now from the right chamber and the upper-left part of the
external field. 
At $t=23.33$~min, instead, the $r^2$ parameter is 0.383, so there is not a collapsed/elongated CS.

%
%
\bibliography{references}
\bibliographystyle{aasjournal}

\end{document}